\documentclass[convention,peer-reviewed]{aesconf}

\graphicspath{{./}{figures/}}

\usepackage[utf8]{inputenc}

\usepackage{microtype}

\usepackage[numbers,square]{natbib}

\usepackage{booktabs}
\usepackage{color}
\usepackage{url}
\usepackage{multirow}
\usepackage{units}
\usepackage{paralist}
\usepackage{graphicx}
\title{Music auto-tagging in the long tail: A few-shot approach}

\author{T.\ Aleksandra Ma}
\author{Alexander Lerch}

\affil{Music Informatics Group, Georgia Institute of Technology}

\correspondence{T.\ Aleksandra Ma and Alexander Lerch}{tengofma@gmail.com, alexander.lerch@gatech.edu}

\lastnames{Ma and Lerch}

\shorttitle{Few-shot music tagging}

\begin{document}

\twocolumn[
\maketitle % MANDATORY!

\begin{onecolabstract}
In the realm of digital music, using tags to efficiently organize and retrieve music from extensive databases is crucial for music catalog owners. Human tagging by experts is labor-intensive but mostly accurate, whereas automatic tagging through supervised learning has approached satisfying accuracy but is restricted to a predefined set of training tags. Few-shot learning offers a viable solution to expand beyond this small set of predefined tags by enabling models to learn from only a few human-provided examples to understand tag meanings and subsequently apply these tags autonomously. 
We propose to integrate few-shot learning methodology into multi-label music auto-tagging by using features from pre-trained models as inputs to {a lightweight linear classifier, also known as a linear probe.} We investigate different popular pre-trained features, as well as different few-shot parametrizations with varying numbers of classes and samples per class. Our experiments demonstrate that a simple model with pre-trained features can achieve performance close to state-of-the-art models while using significantly less training data, such as 20 samples per tag. Additionally, our linear probe performs competitively with leading models when trained on the entire training dataset. The results show that this transfer learning-based few-shot approach could effectively address the issue of automatically assigning long-tail tags with only  limited labeled data. 
\end{onecolabstract}
]

\section{Introduction}
In the rapidly evolving world of digital music, the ability to efficiently organize and retrieve music from expansive databases is crucial for streaming platforms, content creators, and consumers alike. However, mapping tags to musical characteristics is challenging due to the subjectivity inherent in both language and music perception, which varies across individuals and cultures.
Usually, music catalog owners either maintain an in-house annotation team to tag songs in their music repository or use an out-of-the-box music auto-tagging solution. Human tagging, while more flexible and accurate when compared with auto-tagging solutions, is labor-intensive and expensive. Music auto-tagging solutions, on the other hand, are fast and scalable but heavily reliant on a predefined set of (training) tags, as these solutions are commonly trained with supervised learning methods \cite{won2020eval, threeissues}. These supervised methods require a large amount of training data and ground truth label pairs. The reliance on a fixed set of tags  limits their adaptability and personalization to the diverse and dynamic vocabulary of users across different cultures and demographics. For instance, a music branding company curating playlists for a high-end hotel may require a detailed taxonomy of different types of jazz, but most auto-taggers' taxonomies stop at a general classification of jazz. Similarly, wedding music varies significantly across cultures, highlighting the need for more nuanced and customizable tagging systems.

Few-shot learning approaches \cite{wang2020generalizing} present a framework for models to learn new tags from just a few examples, thereby mitigating the need for extensive retraining. This method could enable music catalog owners, such as production music libraries, to save considerable resources on annotation when they need to expand the set of tags, modify their current tag taxonomy, migrate to a new taxonomy, or consolidate two or more taxonomies. An adaptable system thus ensures compatibility across different catalogs and could also enable end users to apply their own, use-case specific taxonomies to their music libraries.

In this study, we propose the utilization of pre-trained audio embeddings as inputs to a simple multi-label, few-shot music auto-tagging model. An ablation study is conducted to examine the impact of the number of classes and the quantity of training samples per class  on model performance to estimate the viability of this approach in  different scenarios. 

In summary, the main contributions of this study are
\begin{itemize}
    \item the introduction of few-shot multi-label classification to the music auto-tagging task,
    \item a comparison of various pre-trained embeddings for music auto-tagging,
    \item a systematic investigation of the effects of number of tag classes and training samples on few-shot classifier performance.

\end{itemize}

The remainder of this paper is structured as follows: Sect. 2 discusses related work in music auto-tagging and few-shot learning. Sect. 3 describes our method, the dataset, the pre-trained features used in our experiments, as well as the methods and experimental setup. Sect. 4 details the results and analysis. Finally, Sect. 5 concludes the paper with a summary of our findings.

\section{Related Work}\label{sec:rel_work}

Music auto-tagging is inherently a multi-label problem \cite{5664796}, meaning that each music track can be associated with multiple tags simultaneously, rather than being limited to a single category. 
Tags span various semantic categories, including genres, instruments, and mood. These tags can range from highly objective, such as "violin," to highly subjective, such as "romantic." Additionally, tags are usually crowd-sourced or user-defined and may present challenges such as ambiguity and subjectivity. There have been efforts that focus on addressing these issues, capturing semantic similarities both directly from audio and from text \cite{hennequin2018audio, epure2019leveraging, epure2020multilingual}.

Initially, efforts to automate the music annotation process focused on leveraging low-level acoustic features, such as zero-crossing rate \cite{mckinneyfeatures}, and mid-level features, such as rhythm \cite{herrera2012}, combined with standard classifiers such as k-Nearest-Neighbors. 
As time progressed, deep learning models have significantly improved auto-tagging performance through carefully designed network architecture \cite{won2020eval, pons2019musicnn, lee2017sample, choi2016automatic}. 
While still requiring data-label pairs, they are much better at capturing complex patterns and representations within the audio data, allowing for better accuracy. With the increase in computing power, large-scale models have been trained on open-access audio datasets \cite{45857} to learn powerful representations that could be used as feature extractors for various downstream tasks in the audio domain \cite{koh2021, hershey2017cnn, lee2017multi}. 
Additionally, there has been an abundance of research aimed at reducing the size and complexity of network architectures by leveraging various pre-trained embeddings \cite{ding2023audio, ding2024embed}. However, previous work in transfer learning for music auto-tagging generally relies on the complete training dataset and may be dependent on specific datasets. Consequently, these methods are often unable to learn an effective classifier from very limited data.

Few-shot learning presents a promising alternative. It refers to a type of machine learning problem where a computer program learns from only a limited number of examples containing supervised information for the target task \cite{wang2020generalizing}. One common formulation of this problem is the \(N\)-way-\(K\)-shot classification, where \(N\)-way represents the number of classes (in our case: tags) the model needs to learn, and \(K\)-shot refers to the number of training samples per class the model uses for learning. In the current literature, few-shot learning is primarily approached in two ways: meta-learning and transfer learning.

Meta-learning approaches aim to simulate the inference process during training by partitioning the training data into support and query sets \cite{snell2017prototypical, sung2018learning, santoro2016meta, finn2017model}.
Transfer learning approaches, on the other hand, involve using pre-trained models as feature extractors and fine-tuning them on the unseen few-shot datasets. Recent research in image classification has demonstrated that transfer learning approaches can yield competitive results comparable to state-of-the-art meta-learning approaches \cite{chen19closerfewshot}, particularly when a diverse set of pre-trained embeddings is combined \cite{chowdhury2021few}. One advantage of transfer learning approaches is that because they are trained on a different set of data, they tend to generalize well and thus can perform well with minimal data, which is particularly important for addressing long-tailed tags with a low number of labeled samples. Therefore, we focus on using pre-trained embeddings in a few-shot setting for multi-label music classification.

\section{Experimental Setup}\label{sec:experiments}

This section details the experimental setup used to evaluate our few-shot classifier design. We describe the methods employed, the dataset and pre-trained features utilized, and the experimental procedures. The primary goal of our experiments is to investigate the performance of few-shot classifiers under various conditions and compare them to state-of-the-art models. 

\begin{figure}[t]

% \includesvg[width=\columnwidth]{experimental_setup}
\includegraphics[width=\columnwidth]{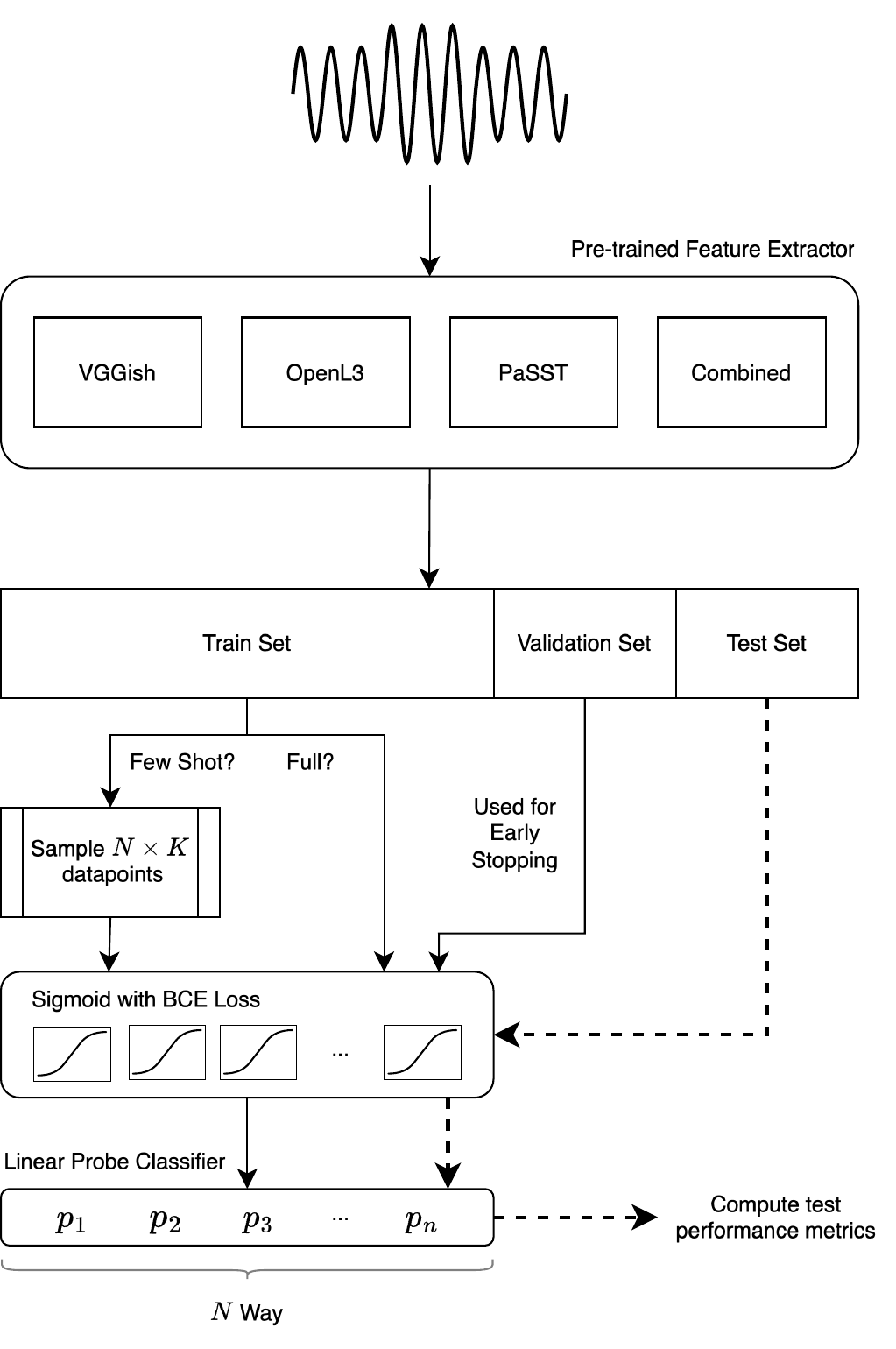}
\caption{
Experimental setup: few-shot linear probes are trained on $N\times K$ number of data points, and full linear probes are trained with all the data in the training set. {Test performance metrics are calculated using probabilities from the Sigmoid activation function on the full test set.}}
\label{fg:exp-setup}

\end{figure}

\subsection{Methods}\label{ssec:methods}

We employ transfer learning approaches by fine-tuning four different types of pre-trained embeddings on a commonly used music auto-tagging dataset through a simple linear probe. {A linear probe is a linear classifier trained on frozen features extracted from a large pre-trained model. It helps to assess how well these embeddings encode task-relevant information} with minimal training. Much of the research in few-shot classification \cite{snell2017prototypical, sung2018learning, santoro2016meta, finn2017model, chen19closerfewshot, chowdhury2021few} frames the target task as a simple multi-class problem; however, music auto-tagging is inherently a multi-label task \cite{5664796}, i.e., multiple tags can be active for the same input. To align our few-shot classifier with the nature of the task, we employ Binary Cross-Entropy loss, coupled with a fully connected layer followed by a Sigmoid activation in our simple linear probe, similar to \citet{wang2021few}. This approach essentially trains a separate logistic regression classifier for each class using a one-vs-rest strategy, thus allowing the concurrent presence of multiple tags in a single sample.

\begin{figure*}
\begin{center}
% \includesvg[width=\textwidth]{mtat_top50_tags_by_split.svg}
\includegraphics[width=\textwidth]{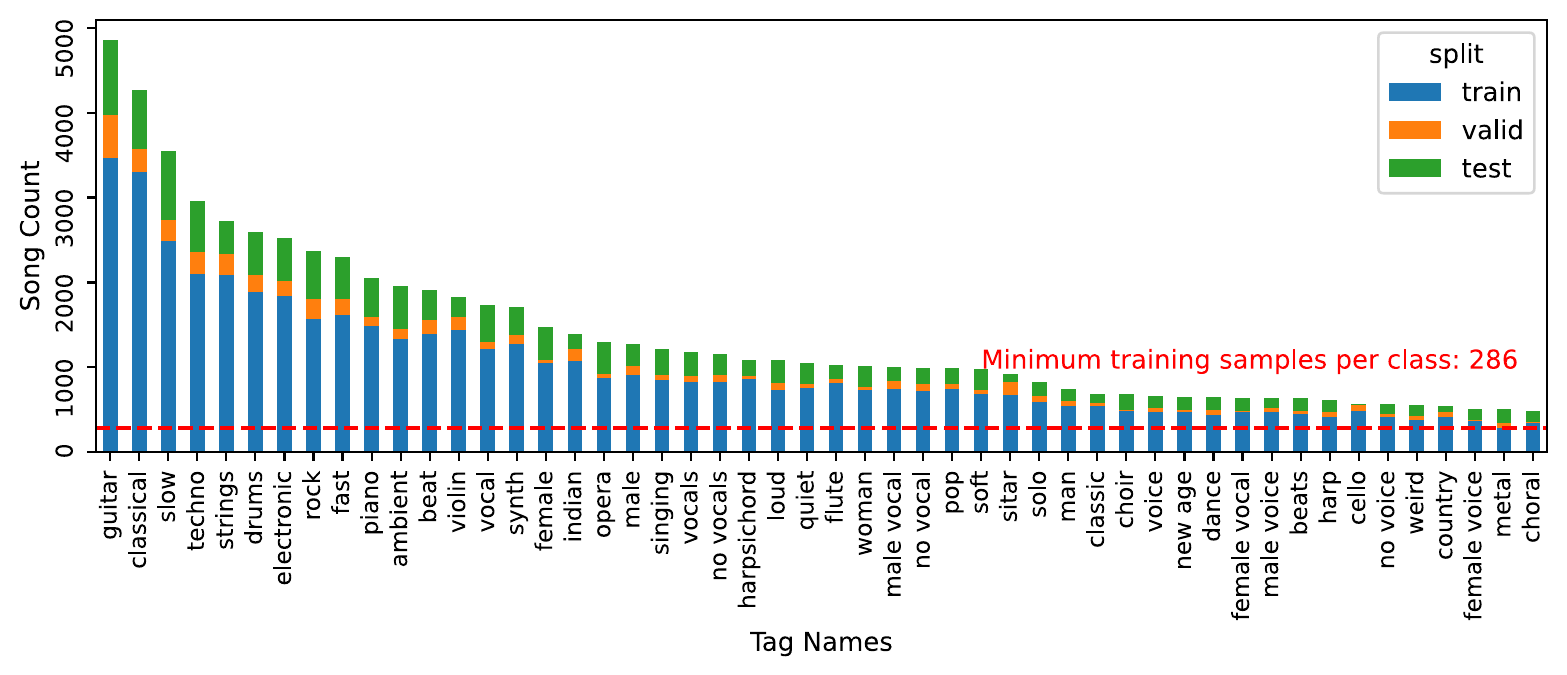}
\caption{Distribution of Top 50 tags of MagnaTagATune by split. }
\label{fg:MTATTag}
\end{center}
\end{figure*}

The experimental setup is detailed in Figure 1. To simulate the few-shot setting, we sample \(N \times K\) ($N$: number of classes, $K$: number of training samples per class) training data points from the training set. The entire validation set is used for early stopping, and the test set is employed to compute the performance metrics of our few-shot classifiers. We use the mean average precision (mAP) and the mean area under the receiver operating characteristics curve (AUC-ROC) to measure the classifier performance to be comparable with previous work\cite{won2020eval, castellon2021}. {Both mAP and AUC-ROC are calculated as the mean of the binary scores for each class, averaged across all classes.} mAP is the mean of the binary average precision score of each class, focusing on how many of the tagged songs are correctly tagged. 
AUC-ROC, on the other hand, is the mean of the binary AUC-ROC scores for each class, focusing more on how many of the songs that do not have a specific tag are wrongfully tagged. 
Using both metrics gives us a comprehensive evaluation of our classifiers. 

\subsection{Data}

The MagnaTagATune dataset \cite{law2009evaluation} is one of the most widely used datasets for automatic music tagging systems. It comprises 25,863 audio clips, each lasting 29.1 seconds. The dataset includes 188 tags, ranging from specific instruments like \textit{violin} or \textit{drums} to broader musical qualities such as \textit{classical} or \textit{jazz}. Most automatic tagging models only utilize the 50 most frequent tags \cite{won2020eval}, discarding the songs not associated with these tags. In our study, we adopt the same split of the top 50 tags as \cite{won2020eval} to ensure our performance metrics are comparable to previous work. Figure 2 shows the distribution of the top 50 tags and their composition in the data split. It is worth noting that even the tag with the least amount of training samples has 286 associated clips. At the same time, the remaining 138 tags that are not in the top 50 tags only have 534 clips in total. The tag with the most training data only has 55 associated clips. This long-tail distribution could also potentially benefit from few-shot approaches.

\subsection{Pre-trained Features}
All pre-trained features are extracted using the whole clip. {We have chosen three pre-trained embeddings –--VGGish, OpenL3, and PaSST--– because they are extracted from models trained on large-scale audio datasets and have demonstrated strong performance in previous studies \cite{hershey2017cnn, arandjelovic2017look,koutini2021efficient}. Each of these embeddings will be introduced in more detail below. }

For these embeddings, multiple frames of embeddings are extracted over the 29.1-second audio clips. These frames are then aggregated into the mean and standard deviation of all frames at each position of the embedding, which results in a vector that is twice as long as its original dimension. This ensures that all four embeddings have the same time resolution, allowing for a fair comparison. 

\subsubsection{VGGish}

VGGish \cite{hershey2017cnn} is a pre-trained model based on the VGG architecture \cite{vgg}, which is known for its deep convolutional layers initially designed for image recognition tasks. It has been adapted for the audio domain by training on a vast collection of YouTube videos. The released VGGish embeddings were trained on \unit[16]{kHz} mono audio and were post-processed through a PCA transformation and quantization to 8 bits. VGGish produces 128-dimensional embeddings. In our setup, the input feature vector is 256-dimensional after aggregation. 
\subsubsection{OpenL3}
OpenL3 \cite{arandjelovic2017look} is a model 
trained through self-supervised learning of audio-visual correspondence in videos. The model we use to extract embeddings from is trained specifically on a music subset of the AudioSet dataset, designed to extract a richer set of audio features. OpenL3 produces 512-dimensional embeddings.  In our setup, the input feature vector is 1024-dimensional after aggregation.

\subsubsection{PaSST}
PaSST \cite{koutini2021efficient}, or Patchwise Self-Supervised Transformer,
is a novel method that optimizes and regularizes transformers on audio spectrograms. The authors propose an efficient training method called Patchout, which also serves as a regularizer, and disentangles the transformer's positional encoding into a time and frequency positional encoding. We use the model variant that has pre-trained time positional encoding for \unit[30]{s} to accommodate for our audio input length. PaSST produces 768-dimensional embeddings.  In our setup, the input feature vector is 1536-dimensional after aggregation.

\subsubsection{Combined Embeddings}
Inspired by \citet{chowdhury2021few}, who demonstrated that building a few-shot classifier on top of multiple pre-trained models results in better classification performance, we combine the three embeddings introduced above. Each embedding is first normalized and aggregated into mean and standard deviation, then concatenated into a 2816-dimensional feature vector.

\subsection{Experiments}

We conducted a series of experiments to comprehensively evaluate our few-shot classifier design, with each research question addressed in the subsections below.
In this section, we refer to models trained with the pre-trained features using the entire training set as "full linear probe" models, and models trained with 
$N \times K$ training data points as "few-shot linear probe" models.

\subsubsection{Exp.~1: Full Linear Probe}

To understand how a simple model architecture using pre-trained features can compare to state-of-the-art models trained from scratch, we trained our simple multi-label neural network with four inputs: VGGish, OpenL3, PaSST, and the concatenation of all three. The resulting four models were trained with the entire training dataset from the original training split. We then compared the performance of the best-performing full linear probe with state-of-the-art auto-tagging models that is trained from audio\cite{won2020eval}, as well as linear probes using other pre-trained models from previous work \cite{castellon2021}.

\subsubsection{Exp.~2: Training Data Efficiency}

Our next research question examines how data efficient our four types of pre-trained features are, e.g., what impact the number of training samples per class has on the classifier performance. We approach this question by examining two metrics: mAP and the correlation coefficient between the trained weights of the few-shot linear probe and the full linear probe. For this experiment, we trained 5 different 50-way few-shot classifiers with each type of pre-trained features, gradually increasing the number of training samples per class ($K$) from 1 to 20. This results in 20 50-way few-shot classifiers. mAP is calculated as described in sect. 3.1. 
Similar to the approach used in \cite{chowdhury2021few}, weight correlation is calculated by taking the L1 norm of the weights at each embedding position. For instance, if we use VGGish to train our linear probe, we have 256 such norms. We then calculate the Pearson correlation coefficient between these norms from the 20-shot classifiers and those from the full linear probe, expecting that a higher correlation indicates performance closer to the full linear probe, the (presumably) best case.

\subsubsection{Exp.~3: Impact of Number of Tags}
In practice, when music catalog owners need to expand their tag vocabulary, they often rely on their internal mapping of tags and music. Thus, the ability of a model to learn effectively from a few samples is crucial. This experiment explores how the number of classes impacts performance and how many songs are needed for our simple few-shot classifier to perform competitively.

We conducted an ablation study by varying the number of classes ($N$) and the number of samples per class ($K$). For each pre-trained embedding, we trained a series of few-shot classifiers, with $N$ and $K$ in the below ranges:
\begin{compactitem}
    \item \(N\) Range: [2, 5, 10, 15, 20, 25, 30, 35, 40, 45, 50]
    \item \(K\) Range: [1, 5, 10, 15, 20]
\end{compactitem}

This resulted in 55 few-shot classifiers for each type of pre-trained embedding, with the number of training data points ranging from 2 ($N=2,\, K=1$) to 1000 ($N=50,\, K=20$). To maintain comparability, classes and training samples were added incrementally. For instance, a 5-way classification model includes the two tags used in a 2-way classification model, and a 20-shot model includes the five samples used in a 5-shot model.

\section{Results and Discussion}\label{sec:results}
This section presents the results from the experiments listed above.

\subsection{Exp.~1: Full Linear Probe}

\begin{table}
\centering

\begin{tabular*}{\columnwidth}{l@{\extracolsep{\fill}}cc}
\toprule
  \textbf{Approach} & \textbf{mAP} & \textbf{AUC-ROC} \\
\midrule
 $\mathrm{LP}_\mathrm{VGGish}$ & 0.42 & 0.90 \\
  $\mathrm{LP}_\mathrm{OpenL3}$ & 0.43 & 0.90 \\
  $\mathrm{LP}_\mathrm{PaSST}$ & 0.45 & 0.91 \\
  $\mathrm{LP}_\mathrm{Combined}$ & \textbf{0.47} & \textbf{0.92} \\
\midrule
 SC-CNN w/ Res \cite{won2020eval} & 0.46 & 0.91 \\
  Probing Jukebox \cite{castellon2021} & 0.41 & \textbf{0.92} \\
\bottomrule
\end{tabular*}
\caption{Performance comparison between our full linear probes, a state-of-the-art auto-tagging model trained from audio, and a probing model using Jukebox embeddings \cite{castellon2021}.}
\label{sota-comp}
\end{table}

\begin{figure}
\centering
% \includesvg[width=.6\columnwidth]{weight_sum_proportion.svg}
\includegraphics[width=.6\columnwidth]{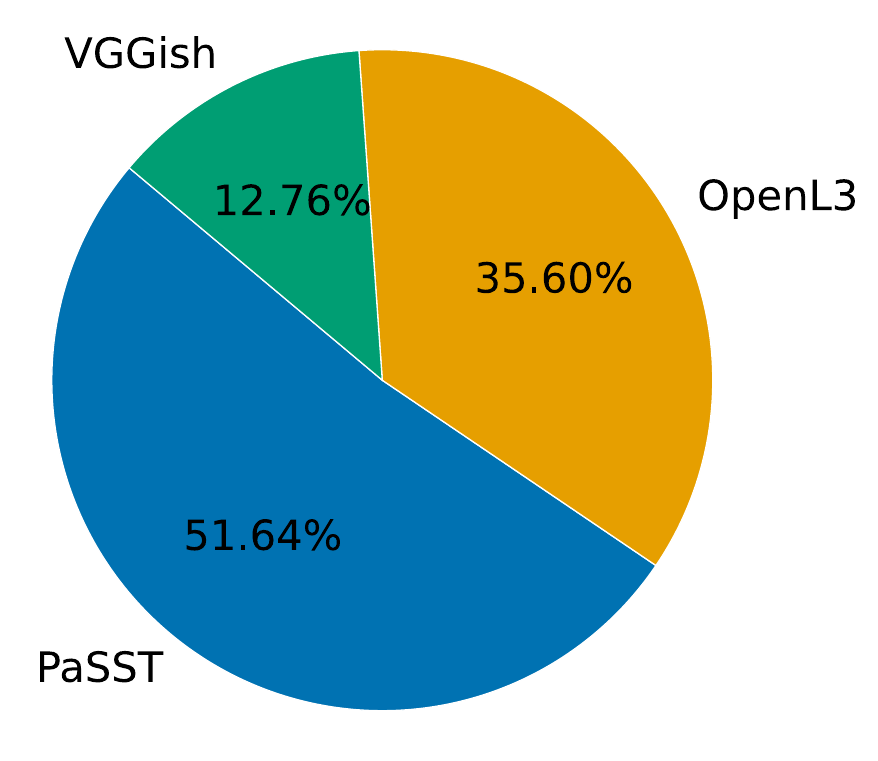}
\caption{Percentage of weight magnitude sum of each pre-trained embeddings in the overall weight magnitude sum of the combined feature. }
\label{fg:combined-prop}
\end{figure}

\begin{figure}
\centering
\includegraphics[width=\columnwidth]{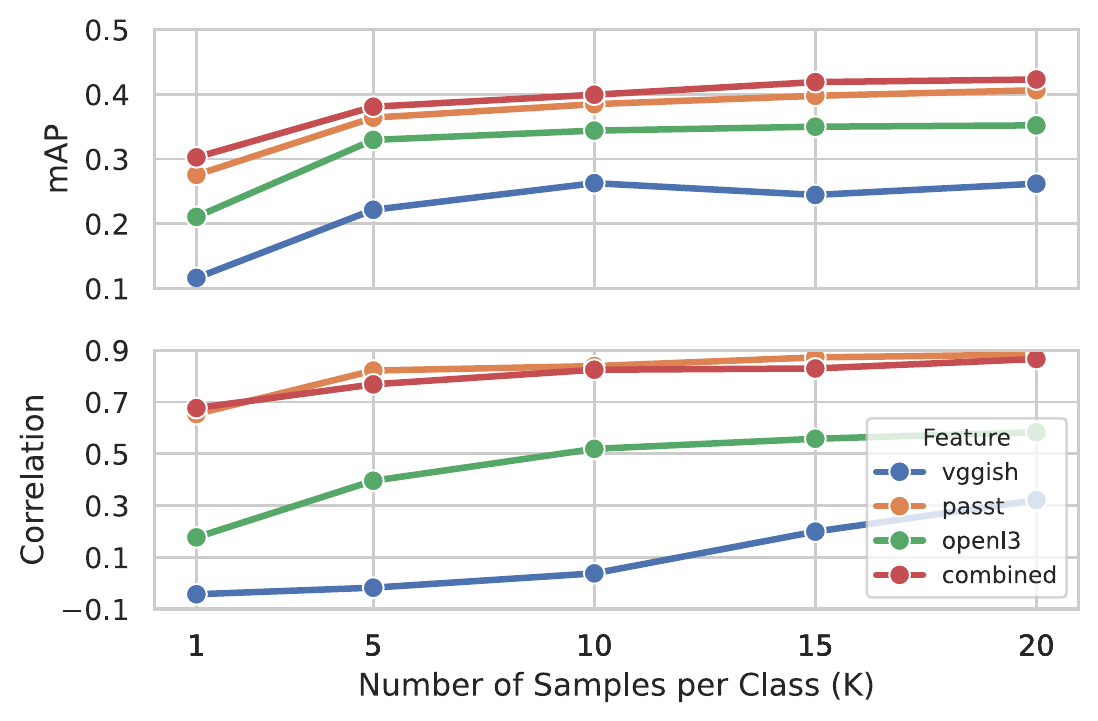}
\caption{Performance comparison of 50-way classifiers dependent on the number of training samples per class: Top: mAP, Bottom: Correlation coefficient between 20-shot probe weights and full probe weights. }
\label{fg:embedding-comp}
\end{figure}

\begin{figure*}
\centering
% \includesvg[width=\textwidth]{performance_change.svg}
\includegraphics[width=\textwidth]{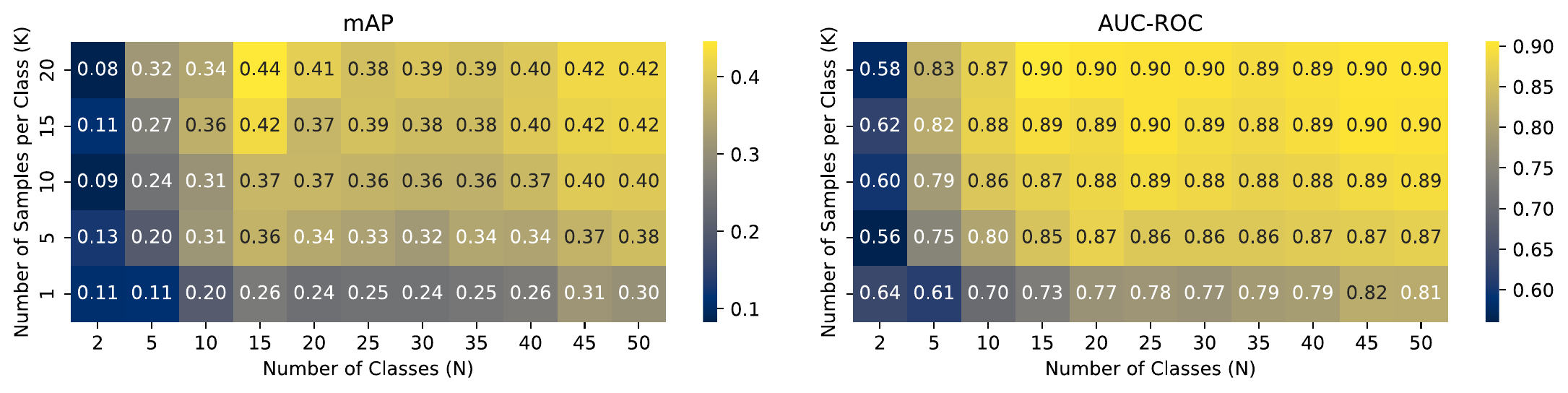}
\caption{Heatmap of how mean average precision (mAP) and area under the receiver operating characteristics curve (AUC-ROC) changes as the number of classes (horiz.) and the number of training samples per class (vert.) increase. Only the best-performing model is shown.}
\label{fg:MTATPerformance}
\end{figure*}

For our first experiment, we report in Table 1 the performance metrics of our linear probes LP$\ast$ trained and evaluated on {29.1-second audio excerpts with} the same data split as proposed by \citet{won2020eval}. We also provide two state-of-the-art performance comparison systems on the top 50 tags of the MagnaTagATune dataset: 
\begin{inparaenum}[(i)]
    \item SC-CNN is a 7-layer CNN with a fully connected layer and  residual connections, trained on 3.69-second audio excerpts, and
    \item Jukebox is a probing classifier with inputs as embeddings extracted by a pre-trained codified audio language model, trained on 29.1-second audio excerpts \cite{castellon2021}. 
\end{inparaenum}

{We observe that our linear probing classifiers perform competitively with both the end-to-end model trained specifically for this task and the linear probe with additional language information encoded.} This indicates that all the pre-trained embeddings used here contain very  relevant information for music tagging and a low-complexity linear classifier is sufficient to achieve results close to the state-of-the-art with such inputs. The PaSST embedding seems to outperform OpenL3 and VGGish embeddings. The better performance of the linear probe classifier with the combined (concatenated) embedding input suggests that our few-shot network learns different aspects from each embedding, and combining them allows these features to complement each other, resulting in higher overall performance metrics.

To investigate the impact of each pre-trained embedding on  $\mathrm{LP}_\mathrm{Combined}$, we calculated the percentage of the weight magnitude of each single embedding in the weight magnitude of the combined embedding. Figure 3 shows that the PaSST embeddings have the highest impact, followed by OpenL3 and VGGish.

\subsection{Exp.~2: Training Data Efficiency}
The second experiment investigates the data efficiency of each pre-trained features through observing the impact of the number of training samples per class for a 50-way classifier.

The top plot in Fig. 4 shows the mAP of the linear probing classifiers with different types of pre-trained embeddings as the number of samples per class increases. Unsurprisingly, the mAP increases in most cases with increasing number of training samples, but increasing the number of training samples per class beyond 10 only yields small performance gains. $\mathrm{LP}_\mathrm{PaSST}$ tends to consistently show the best performance across all training set sizes, followed by $\mathrm{LP}_\mathrm{OpenL3}$ and $\mathrm{LP}_\mathrm{VGGish}$. Using combined embeddings as input yields better metrics than any of the single embeddings. 
Figure 4 bottom shows the weight correlation between the linear probe classifiers trained with different training set sizes and the weights of the full linear probe that was trained with all available training data. Assuming that the latter represents the best-case outcome, the plot shows increasing correlation with increasing number of training samples. $\mathrm{LP}_\mathrm{PaSST}$ and $\mathrm{LP}_\mathrm{Combined}$, in particular, show a high correlation starting from a low number of training samples, indicating highly efficient training data utilization.

\subsection{Exp.~3: Impact of Number of Tags}
While the results presented above focused on the performance in a 50-way scenario, i.e., a classification into 50 classes, in this experiment we evaluate how the number of classes (or labels) influences the performance. As the number of target classes can vary dependent on the specific task, understanding the impact of this can be crucial in a real-world scenario.

Figure 5 depicts the changes in mAP and AUC-ROC as the number of samples per class increases for classifiers with varying numbers of classes. The metrics shown represent the best performing pre-trained features discussed in the previous subsection. As reported in Exp.~2, performance increases as we increase the number of training samples per class $K$ for the classifiers with the same number of classes $N$ (the results of Exp.~2 can be revisited in the right column of the matrix). 
On the other hand, as we increase the number of classes while keeping the number of samples per class constant, performance tends to be stable except for very low N's and K's. Even though increasing the number of classes adds negative examples for the previous classes in the training set, it seems like simple binary classifiers are not dramatically impacted by this training set imbalance. 

It is important to note that all tags in the test set always have the same amount of positive and negative samples in all our experiments because we use the entire test set to calculate performance metrics. If we constrained the test set to include only songs with at least one associated tag, it would inflate the AUC-ROC with increasing class number. This is because AUC-ROC involves measuring the false positive rate (FPR), calculated by dividing the number of false positives by the sum of true negatives and false positives. With the constraint, moving from a lower number of classes to a higher number of classes increases the number of true negatives in the test set, thus exaggerating the AUC-ROC.

\section{Conclusion}\label{sec:conclusion}

In this paper, we explored the potential of few-shot learning for multi-label music auto-tagging by training a linear probe using pre-trained embeddings. The pre-trained embeddings we experimented with are VGGish, OpenL3, PaSST, and all three combined. Our results demonstrate that combining embeddings from different pre-trained models can yield both performance on par with state-of-the-art models and improved training data efficiency. We showed that using pre-trained embeddings with linear probing is also a very suitable approach for few-shot settings and systematically evaluated the performance loss introduced by the few-shot scenario. Our results emphasize that few-shot learning utilizing pre-trained embeddings can be an effective strategy for music tagging and a viable solution for long-tail tags with limited labeled samples.

To our knowledge, this paper is the first attempt to investigate multi-label few-shot learning for music auto-tagging. We believe this approach can effectively handle situations where only a limited amount of data is available. In the future, we plan to explore methods for incorporating  novel classes learned from a few-shot classifier into existing auto-tagging models, in order to build a truly customizable auto-tagging system.

\bibliographystyle{jaes}

% Reference to bibliography file.
\bibliography{refs}

\end{document}